\documentclass[twocolumn,prl,showpacs,preprintnumbers,amsmath,amssymb]{revtex4}

\usepackage{graphicx}
\usepackage{dcolumn}
\usepackage{bm}

\begin{document}
\preprint{XXX/XXX-XXX}
\title{Direct Optical Excitation of a Fullerene-Incarcerated Metal Ion}

\author{Mark A G Jones}\affiliation{QIPIRC, Department of Materials, University of Oxford, Parks Road, Oxford, Oxon OX1 3PH, United Kingdom}\email{mark.jones@materials.ox.ac.uk}
\author{R A Taylor}
\author{A Ardavan}\affiliation{Clarendon Laboratory, Department of Physics, University of Oxford, Parks Road, Oxford, Oxon OX1 3PU, United Kingdom}
\author{K Porfyrakis}\author{G A D Briggs}\affiliation{QIPIRC, Department of Materials, University of Oxford, Parks Road, Oxford, Oxon OX1 3PH, United Kingdom}

\date{\today}

\begin{abstract}
The endohedral fullerene $\mathrm{Er_3N@C_{80}}$ shows characteristic $\mathrm{1.5\;\mu m}$ photoluminescence at cryogenic temperatures associated with radiative relaxation from the crystal-field split $\mathrm{Er^{3+}}$ $^4\mathrm{I}_{13/2}$ manifold to the $^4\mathrm{I}_{15/2}$ manifold. Previous observations of this luminescence were carried out by photoexcitation of the fullerene cage states leading to relaxation via the ionic states. We present direct non-cage-mediated optical interaction with the erbium ion. We have used this interaction to complete a photoluminescence-excitation map of the $\mathrm{Er^{3+}}$ $^4I_{13/2}$ manifold. This ability to interact directly with the states of an incarcerated ion suggests the possibility of coherently manipulating fullerene qubit states with light.
\end{abstract}

\maketitle

The endohedral environment of atoms and ions incarcerated in fullerene cages screens the guest species from external physical and chemical influences \cite{Har02}. This screening interferes with the ability to study and control the endohedral species. Spectroscopic data can be used in conjunction with parametric modelling to elucidate the local electronic environment \cite{Rei83}, but strong optical absorption in the visible and near infrared region associated with the fullerene cage \cite{Aji90} constrains the use of optical emission and absorption techniques.

\begin{figure}
\includegraphics[scale=0.25]{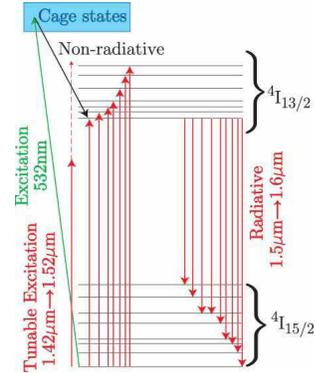}
\caption{Level scheme of the Er$^{3+}$ ion in Er$_3$N@C$_{80}$, showing excitation and relaxation processes for $\mathrm{532\,nm}$ excitation and for $\mathrm{1.5\,\mu m}$ excitation. \label{levels}}
\end{figure}

Fluorescent emission has been measured from $\mathrm{Er^{3+}}$ ions in both $\mathrm{Er_2@C_{82}}$ \cite{Mac97} and the tri-nitride-template fullerenes $\mathrm{Er}_x\mathrm{Sc}_{3-x}\mathrm{N}\mathrm{@C_{80}}{[x=1,2,3]}$\cite{Mac01}. In both cases, excitation is induced in the fullerene cage electronic states by laser pumping at a wavelength that is absorbed by the fullerene cage. The excitated states relax nonradiatively and incoherently to the ionic states with subsequent luminescencent relaxation between the $\mathrm{Er^3+}$ ion $^4I_{13/2}$ and $^4I_{15/2}$ manifolds at a wavelength longer than the cage absorption edge. At cryogenic temperatures, the luminescence is resolved into discrete peaks corresponding to transitions between the crystal-field-split components of each of these manifolds.  In both materials, the only transitions visible in these experiments have been from the lowest two sublevels of the upper manifold to the 8 sublevels of the lower manifold (Figure \ref{levels}). Rapid nonradiative relaxation within the upper manifold precludes the observation of radiative transitions from other levels within the upper manifold, preventing the full level structure of the upper manifold being observed in the emission spectrum.

\begin{figure}

\includegraphics[width=0.4\hsize, scale=0.5]{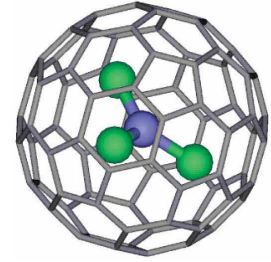}
\caption{Er$_3$N@C$_{80}$ structural model. The enclosed tri-erbium nitride template aligns with a cage threefold axis.\label{model}}
\end{figure}

In this Letter, we report the full level structure of the upper manifold measured by using photoluminescence-excitation (PLE) spectroscopy to directly excite the individual ground-state to upper-manifold transitions. The energies and linewidths of these transitions corroborate the narrow linewidths observed in this system and provide further evidence for the highly ionic character of the endohedral erbium. More importantly, this technique demonstrates direct coherent optical interaction with an endohedral ion. Fullerene-based quantum computing schemes \cite{Ard03} which employ an endohedral spin qubit will require single-spin readout. Optical readout methods show promise and have been implemented in the diamond N-V system \cite{Jel04}. Direct optical excitation resulting in luminescence may provide a pathway to efficient and effective readout of such schemes\cite{Ben06}.

As the cage exhibits transparency for wavelengths longer than $\mathrm{1\,\mu m}$ \cite{Kik94}, it is in principle possible to coherently excite the upper manifold levels of the endohedral ion directly, using a tunable laser (Figure \ref{levels}). When the excitation laser is resonant with one of the $\mathrm{Er^{3+}}$ $\mathrm{^4I_{15/2}(1)\rightarrow^4I_{13/2}(1\dots7)}$ transitions, excitation into the upper manifold will occur. When the excitation laser is off-resonance, no such process is allowed. By observing the luminescence as a function of excitation wavelength, the transitions from the ground state of the ion to the upper manifold levels may be identified and their relative transition strengths recorded.

This technique not only allows the probing and mapping of states whose luminescence is inhibited by nonradiative decay, but also permits the selective excitation of individual ionic states. This is not possible by excitation via the cage, as the nonradiative relaxation process populates multiple ionic states statistically and incoherently. Such control is a prerequisiste for fullerene quantum optics or optical fullerene qubit manipulation.

$\mathrm{Er_3N@C_{80}}$ powder manufactured by Luna Innovations Inc.\ using the arc-discharge method \cite{Ste99} was passed through high-performance liquid chromatrography (HPLC) apparatus to ensure its purity before use. UV-Vis-NIR absorbtion spectra were acquired to confirm the identity of the collected fraction \cite{Kik94}.

A saturated solution sample of $\mathrm{Er_3N@C_{80}}$ dissolved in carbon disulphide ($\mathrm{CS_2}$) was sealed under vacuum in a 4mm OD quartz tube and suspended in an Oxford Instruments CF442 cryostat arranged for fluorescence measurement. Fluorescence spectra perpendicular to the excitation beam path were acquired with a Princeton Instruments OMA-V-1024LN InGaAs array detector. Dispersion of the fluorescence was achieved by imaging the sample onto the entrance slit of a Spex 881 triple monochromator. For all experiments, the sample was cooled to $\mathrm{5\;K}$ with continuous helium flow.

\begin{figure}
\includegraphics[scale=0.35]{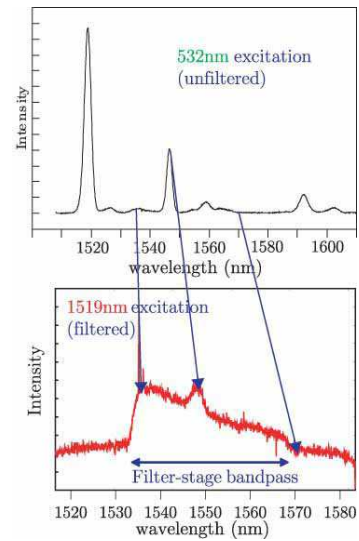}
\caption{Observation of direct optical interaction with an incarcerated ion. Upper spectrum is the eight-line fluorescence spectrum of $\mathrm{Er_3N@C_{80}}$ at $\mathrm{5\,K}$ excited by $\mathrm{532\,nm}$ laser radiation, as previously reported \cite{Mac01}. The lower spectrum is the fluorescence spectrum, in our apparatus, of $\mathrm{Er_3N@C_{80}}$ under excitation with $\mathrm{1519\,nm}$ laser radiation. The filter bandpass permits observation of the second-strongest peak in the fluorescence spectrum while the laser is tuned to the wavelength of the strongest peak. The asymmetric background is due to residual laser scatter.\label{firstlight}}
\end{figure}

Initial alignment of the system was performed using $\mathrm{532\,nm}$ green emission from a CW DPSS laser ($\mathrm{5\,mW}$), a wavelength at which the cage absorbs strongly \cite{Kik94}. The green laser was applied to the sample to observe the 8-line spectrum around $\mathrm{1.55\;\mu m}$ \cite{Mac01}  (Figure \ref{firstlight}).

Infrared excitation was provided by a Santec TSL-210V tunable laser diode operating between $\mathrm{1.420\;\mu m}$ and $\mathrm{1.520\;\mu m}$ with maximum operating power of $\mathrm{18\;mW}$ and a linewidth of $<\mathrm{1\,MHz}$. To demonstrate the principle of direct excitation of the ionic energy levels, laser excitation was applied at the wavelength of the strongest emission line in the luminescence spectrum ($\mathrm{^4I_{13/2}(1)\leftrightarrow^4I_{15/2}(1)}$, $\mathrm{1519\,nm}$) and the luminescence of the second-strongest line ($\mathrm{^4I_{13/2}(1)\rightarrow^4I_{15/2}(4)}$, $\mathrm{1546\,nm}$ was observed (Figure \ref{levels}). The laser was tuned until the scattered laser light was observed at the same detector pixel as the centre of the strongest emission line. As the excitation and observed light are close in wavelength, and the luminescence is weak compared to the scattered laser light, the monochromator filter stage was introduced and adjusted in both centre wavelength and bandpass until only a small range of wavelengths surrounding the region of interest were passed to the dispersive stage for recording.

Figure \ref{firstlight} shows the peak observed in the region of interest when exciting with $\mathrm{10\;mW}$ of light at this wavelength. The integration time was $1\,\mathrm{minute}$. The background observed was attributed to a combination of laser spontaneous background emission and residual laser scatter. 
When the experiment was repeated with the laser detuned by $\mathrm{2\;nm}$, this peak was no longer observed, while the background remained.

Having demonstrated that optical direct interaction is indeed possible with this ion, an attempt was made to excite into the higher components of the $^4I_{13/2}$ manifold. Each principal emission line was in turn observed through the filter passband and the experiment repeated, stepping the laser over an appropriate range in $\mathrm{0.1\;nm}$ steps. The spectral region of interest was integrated and the background subtracted to produce an excitation spectrum of the $^4I_{13/2}$ manifold.

\begin{figure}
\includegraphics[scale=0.25]{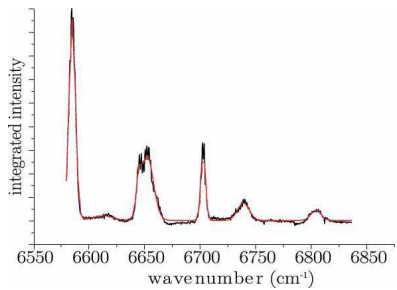}
\caption{Integrated luminescence intensity with excitation photon wavenumber, and seven-Gaussian fit.\label{energy}}
\end{figure}

At $\mathrm{5\,K}$, only the lowest level of the ground manifold is significantly populated, so all excitation peaks visible in the excitation spectrum of the second-strongest photoluminescence peak arise from the 7 $\mathrm{^4I_{15/2}(1)\rightarrow^4I_{13/2}(\mathit{n})}$ transitions. This spectrum was then transformed into energy space for peak fitting (Figure \ref{energy}).

The well-resolved and sharp nature of the excitation lines observed indicate that the luminescent centre retains its $\mathrm{Er^{3+}}$ ionic character across the $\mathrm{^4I_{13/2}}$ manifold. The peak energies and their linewidths and intensities are listed in Table \ref{peaks}. Since the laser linewidth is much smaller than that of the peak, the excitation spectrum gives an accurate spectral profile for the transition. The comparable linewidth and wavelength of the excitation and emission of the $\mathrm{^4I_{15/2}(1)\leftrightarrow^4I_{13/2}(\mathrm{1})}$ line confirm that the interaction is directly with the ionic state, rather than via a molecular or cage orbital. The lines of the excitation spectrum are of Gaussian, rather than Lorentzian or Voigt, profile, indicating that the transitions are inhomogeneously broadened. This suggests that although the ion is screened inside the fullerene cage, local crystalline variation in the polycrystalline $\mathrm{CS_2}$ solvent distorts the cage, causes variations in the local crystal field and leads to site-to-site variation in the fullerene spectral profile.

\begin{table}
\caption{\label{peaks} PLE Peaks}
\begin{tabular}{l|l|l}
Wavenumber ($\mathrm{cm^{-1}}$)&Width ($\mathrm{cm^{-1}})$&Rel. Intensity\\
6584.2&	5.8&	1.000\\
6611.6&	18.0&	0.072\\
6644.0&	3.6&	0.096\\
6651.5&	11.1&	0.625\\
6701.9&	4.0&	0.245\\
6738.1&	10.4&	0.157\\
6803.4&	8.8&	0.079\\
\end{tabular}
\end{table}

\begin{figure}
\includegraphics[scale=0.18]{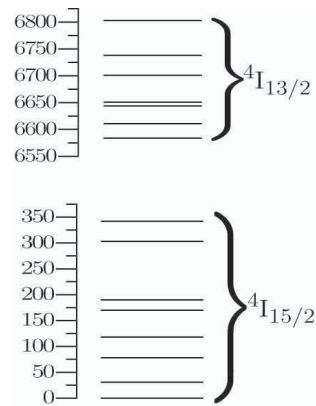}
\caption{Map of observed states in $\mathrm{Er_3N@C_{80}}$, using data from the PLE map and the previous luminescence  report\cite{Mac01}. \label{levels_nums}}
\end{figure}

Combining the PLE map with analysis of data from fluorescence spectrum \cite{Mac01} permits contruction of the ionic level structure of the $\mathrm{Er_3N@C_{80}}$ system (Figure \ref{levels_nums}).
The splitting of the upper manifold is around $250\,\mathrm{cm^{-1}}$, comparable to the splitting of the lower manifold ($350\,\mathrm{cm^{-1}}$), and comparable to the splittings observed in ionic hosts. This, together with the narrow inhomogeneous linewidth, confirms that the $\mathrm{Er}^3+$ ion in this molecule retains significant ionic character despite the incorporation into the tri-nitride cluster and the fullerene cage.

We have thus optically interacted directly, selectively and coherently with an ion encapsulated within a fullerene cage. This direct excitation spectroscopy has enabled us to produce a map of the first excited manifold of the encapsulated endohedral ion. The results of this technique may, through crystal-field analysis, shed light on the local environment experienced by the endohedral system, which, due to the presence of the lanthanide electronic structure, is difficult to model via traditional molecular modelling codes.

By exploiting transparent regions in the cage absorption to directly drive ionic transitions, this gives an additional tool with which to manipulate and probe the quantum state of endohedral species.

\begin{acknowledgments}
This research is part of the QIP IRC www.qipirc.org (GR/S82176/01). This work also has been supported by Hewlett-Packard Co. through the DARPA Slow Light programme, contract FA9550-05-C-0017. GADB is supported by an EPSRC Professorial Research Fellowship (GR/S15808/01). AA is supported by the Royal Society. $\mathrm{Er_3N@C_{80}}$ was supplied by Luna Innovations, Blacksburg, VA, USA. We thank Michael Reid, Robert Davies, Ian Walmsley and Ray Beausoleil for stimulating discussions.
\end{acknowledgments}

\end{document}